\documentclass[sigconf, natbib=false, nonacm]{acmart}

\usepackage[datamodel=acmdatamodel, style=acmnumeric, sorting=none, backend=biber]{biblatex}
\usepackage{amsmath, amsfonts}

\usepackage{orcidlink}

\usepackage{graphicx}
\usepackage{subcaption}

\usepackage[inline]{enumitem}
\setlist*[itemize]{labelindent=10pt, itemindent=0pt, leftmargin=*}

\usepackage{booktabs}
\usepackage{multirow}
\usepackage{tabularx}

\usepackage{pgfplots}
\pgfplotsset{compat=1.18}
\usepgfplotslibrary{statistics}

\usepackage[nameinlink]{cleveref}

\usepackage{csquotes}
\usepackage{textcomp}

\usepackage{balance}

\setcopyright{rightsretained}
\copyrightyear{2025}

\begin{document}

\title[Demonstrators for Industrial CPS Research]{Demonstrators for Industrial Cyber-Physical System Research: A Requirements Hierarchy Driven by Software-Intensive Design}

\author{Uraz {Odyurt}}
\orcid{0000-0003-1094-0234}
\affiliation{%
	\institution{Faculty of Engineering Technology, University of Twente}
	\city{Enschede}
	\country{The Netherlands}}
\email{u.odyurt@utwente.nl}

\author{Richard {Loendersloot}}
\orcid{0000-0002-1113-8203}
\affiliation{%
	\institution{Faculty of Engineering Technology, University of Twente}
	\city{Enschede}
	\country{The Netherlands}}
\email{r.loendersloot@utwente.nl}

\author{Tiedo {Tinga}}
\orcid{0000-0001-6600-5099}
\affiliation{%
	\institution{Faculty of Engineering Technology, University of Twente}
	\city{Enschede}
	\country{The Netherlands}}
\email{t.tinga@utwente.nl}

\renewcommand{\shortauthors}{U. Odyurt et al.}

\begin{abstract}
One of the challenges apparent in the organisation of research projects is the uncertainties around the subject of demonstrators. A precise and detailed elicitation of the coverage for project demonstrators is often an afterthought and not sufficiently detailed during proposal writing. This practice leads to continuous confusion and a mismatch between targeted and achievable demonstration of results, hindering progress. The reliance on the Technology Readiness Level (TRL) scale as a loose descriptor does not help either. We propose a demonstrator requirements elaboration framework aiming to evaluate the feasibility of targeted demonstrations, making realistic adjustments, and assist in describing requirements. In doing so, we define 5 hierarchical levels of demonstration, clearly connected to expectations, e.g., work package interaction, and also connected to the project's industrial use-cases. The considered application scope in this paper is the domain of software-intensive systems and industrial cyber-physical systems. A complete validation is not accessible, as it would require application of our framework at the start of a project and observing the results at the end, taking 4--5 years. Nonetheless, we have applied it to two research projects from our portfolio, one at the early and another at the final stages, revealing its effectiveness.

\end{abstract}

%


\maketitle


\section{Introduction}
\label{sec:introduction}
Research projects often include \enquote{demonstrators} as deliverables. It is often assumed that a demonstrator is inherently known to project members, with no structure or definitive description provided as part of the project proposal. Another valid reason for the absence of such a description is the initial level of uncertainty specific to research projects. After all, a research project is supposed to be high-level by definition. Questions such as: \enquote{What is there to be demonstrated?}, \enquote{How much of it can be demonstrated?}, \enquote{Given the a priori knowledge limitations resulting from unknowns in a research project, which level of demonstration is feasible?}, are familiar to most researchers. Most projects try to solve this by relying on the Technology Readiness Levels (TRLs)~\cite{Mankins:2004:TRL}, which proves to be inadequate as it does not account for a demonstrator's coverage. We discuss this in \Cref{subsec:comparison_to_trl_scale}.

The vague and undefined nature of a \enquote{demonstrator} leads to misaligned expectations, excessive delays, or simply failing to provide a viable demonstrator. This is an especially apparent shortcoming with research projects in collaboration with the industry. Research projects in natural science domains with an experimental focus, e.g., experimental high-energy physics, and more obviously, projects for various engineering domains, e.g., computer science, can be mentioned.

A key aspect that is often left implicit in the context of demonstrators is the distinction between two layers of requirements. \emph{The first layer concerns the requirements governing the formulation and alignment of demonstrators at the project-level}, i.e., what is to be demonstrated and to what extent. The second layer concerns the more granular requirements associated with each individual demonstrator's implementation, which are inherently dependent on the underlying system design, particularly in the case of software-intensive systems. In this work, we focus on the first layer, as it is foundational for reducing ambiguity and aligning expectations early in the project lifecycle.

\paragraph*{Contribution}
We propose a systematic approach for a clear definition of requirements, clarifying a research project's demonstrator(s) ambitions at the project-level. We take inspiration from common research project structures, as well as the TRL scale. In this particular paper, we consider a software-intensive system as the deliverable, which would clarify additional unknowns. We provide:
\begin{itemize}
	\item A structured and practical typology of demonstration levels, suitable for research projects.
	\item A systematic and reusable framework designed to reveal misalignments and/or missing detail crucial to the composition of demonstrators.
	\item Application of our method (framework) to two research projects in our portfolio.
\end{itemize}

\paragraph*{Scope of research projects}
The role of software is ever-increasing in industrial Cyber-Physical Systems (CPS), to an extent that we can consider modern instances as \emph{Software-Intensive Systems (SIS)}~\cite{Hölzl:2008:ESIA, Oquendo:2016:SACE}. This designation is backed by the fact that analytic solutions intended for industrial CPS, especially the likes of solutions providing capabilities such as Fault Detection and Isolation (FDI)~\cite{Hwang:2010:SFDI}, Fault Detection and Diagnosis (FDD)~\cite{Isermann:2005:MFDD}, anomaly detection/identification~\cite{Chandola:2009:ADAS, Ibidunmoye:2015:PADB}, and predictive maintenance, are fully based on software algorithms and data. In this paper, we focus on research projects that consider use-cases from the domain of industrial CPS, with an expected outcome in the context of SIS. Further evaluations with more diverse projects could lead to adjustments, making our method suitable for a broader scope.

Following this introduction, \Cref{sec:background} covers fundamental topics and \Cref{sec:demonstrator_taxonomy} presents our proposed demonstrator typology. Our designed framework as the method is covered in \Cref{sec:framework}, with example cases analysed in \Cref{sec:discussion}. Related works and concluding remarks are given in \Cref{sec:related_work} and \Cref{sec:conclusion}, respectively.

\section{Background}
\label{sec:background}
A few fundamental topics are to be elaborated.

\subsection{TRL scale}
The Technology Readiness Level (TRL) scale in its original form, was first formalised by NASA~\cite{Mankins:2004:TRL}. The scale is intended as a framework for systematic assessment of technology maturity during its development. Consisting of 9 levels from basic research (TRL~1) to fully operational (TRL~9), it was primarily focused on hardware for aerospace technologies. These levels are defined as (taken from~\cite{Mankins:2004:TRL}):
\begin{itemize}
	\item TRL~1: Basic principles observed and reported
	\item TRL~2: Technology concept and/or application formulated
	\item TRL~3: Analytical and experimental critical function and/or characteristic proof-of-concept
	\item TRL~4: Component and/or breadboard validation in laboratory environment
	\item TRL~5: Component and/or breadboard validation in relevant environment
	\item TRL~6: System/subsystem model or prototype demonstration in a relevant environment (ground or space)
	\item TRL~7: System prototype demonstration in a space environment
	\item TRL~8: Actual system completed and \enquote{flight qualified} through test and demonstration (ground or space)
	\item TRL~9: Actual system \enquote{flight proven} through successful mission operations
\end{itemize}

The TRL scale can act as a common language to track, assess and communicate technology maturity across involved organisations or even different projects.

\subsection{Adapted TRLs for SIS}
Modern systems portray an increasing reliance on software components to achieve the core functionality. Considering the differences between hardware development and software development, reinterpretation of TRLs to better reflect the unique characteristics of software development and deployment is necessary. An adapted version of the TRL scale, focusing on software-intensive systems is a sensible update to have. The key motivating differences are:
\begin{itemize}
	\item Software development lacks physical instantiation stages, making traditional TRL interpretations less applicable.
	\item Software is typically developed using agile or incremental methods, requiring readiness to reflect iterative validation.
	\item The maturity of software is often more dependent on its interaction with other systems than on standalone functionality, i.e., there is a clear distinction between individual functionality and integrated functionality.
	\item Software must be tested under realistic conditions, especially in safety-critical or real-time environments, to achieve robustness.
\end{itemize}

Given the above differences and considering the original TRL scale, \Cref{tab:trl_levels} provides the adapted TRL scale definitions for software-intensive systems. Software-specific keywords and concepts are introduced in blue text for legibility and ease of comparison.
\begin{table*}[htbp]
	\centering
	\caption{Adapted TRL scale definitions to accommodate key differences present in software development and software-intensive systems. Adapted text is highlighted in blue font.}
	\label{tab:trl_levels}
	\begin{tabularx}{\textwidth}{@{}lXX@{}}
		\toprule
		\textbf{Level} & 
		\textbf{Original definition} & 
		\textbf{Adapted definition} \\
		\midrule
		TRL~1 	& Basic principles observed and reported & Basic principles or \textcolor{blue}{algorithms} identified and described \\
		TRL~2 	& Technology concept and/or application formulated & \textcolor{blue}{Software} concept and intended application formulated \\
		TRL~3 	& Analytical and experimental critical function and/or characteristic proof-of-concept & Proof-of-concept \textcolor{blue}{algorithms} developed and demonstrated in analytical or \textcolor{blue}{simulation environments} \\
		TRL~4 	& Component and/or breadboard validation in laboratory environment & Individual \textcolor{blue}{software components} prototyped and tested in \textcolor{blue}{controlled lab conditions} \\
		TRL~5 	& Component and/or breadboard validation in relevant environment & \textcolor{blue}{Software modules integrated} and tested with \textcolor{blue}{representative data and interfaces} in a \textcolor{blue}{simulated or partially relevant environment} \\
		TRL~6 	& System/subsystem model or prototype demonstration in a relevant environment (ground or space) & \textcolor{blue}{Prototype software} integrated with relevant subsystems and tested in a relevant environment, e.g., \textcolor{blue}{hardware-in-the-loop} \\
		TRL~7 	& System prototype demonstration in a space environment & \textcolor{blue}{Integrated software prototype} demonstrated in an \textcolor{blue}{operational or high-fidelity simulated environment} \\
		TRL~8 	& Actual system completed and \enquote{flight qualified} through test and demonstration (ground or space) & \textcolor{blue}{Fully developed software system} qualified through rigorous testing under expected operational conditions \\
		TRL~9 	& Actual system \enquote{flight proven} through successful mission operations & \textcolor{blue}{Software system validated} through successful operational use in the \textcolor{blue}{target environment} \\
		\bottomrule
	\end{tabularx}
\end{table*}

\subsection{Project context and stakeholders}
Research projects, especially those co-funded by public-private partnerships or national/international agencies, often bring together a diverse range of stakeholders. These typically include academic institutions, industrial collaborators, applied research organisations, and funding or oversight bodies. Each actor contributes unique capabilities, objectives, and expectations to the project. While this diversity enables innovation, it can also complicate coordination, especially regarding shared outputs like demonstrators.

\subsubsection{Stakeholders and building blocks}
The typical structure of such projects revolves around \emph{Work Packages (WPs)}, which are thematic divisions that organise tasks, responsibilities, and deliverables. WPs may focus on core research, integration, validation, dissemination, or management. The interplay between WPs reflects both the workflow and the interdisciplinary nature of the project. Accordingly, common actors in research projects are,
\begin{itemize}
	\item \emph{Academic stakeholders} often lead fundamental research, algorithm development, or methodology design.
	\item \emph{Industrial partners} usually provide real-world use-cases, validation platforms, or deployment opportunities.
	\item \emph{Applied research institutes} act as bridges, ensuring that scientific advances translate into implementable technology.
	\item \emph{Funding/oversight bodies} ensure alignment with broader strategic goals and assess milestone completion.
\end{itemize}

This landscape of actors leads to multiple perceptions of what a demonstrator should achieve. For instance, academics may consider a functional proof-of-concept sufficient, while industry partners may expect operational integration and robustness. Without a shared typology/set of definitions, mismatches in expectations are common, often becoming apparent too late in the project timeline.

Demonstrator-related misalignments typically map to misunderstood, underspecified, or unspecified requirements. Especially in software-intensive projects, the absence of clearly defined interfaces, data formats, or operational contexts between WPs leads to friction during integration phases. Therefore, Requirements Engineering (RE) practices in research contexts should explicitly address targeted demonstrator scope and maturity.

\subsubsection{WP dependencies}
Considering the four major dependencies between software components, i.e., data dependency, control dependency, functional dependency, and temporal dependency, the most relevant dependencies for WPs can be considered as data and temporal. While we cannot dismiss control and functional dependencies, especially given the possibility of per WP software converging into a unified platform, data and temporal dependencies stand out. Incorporation of control or functional dependencies is less common as these do not align with having separate WPs. Most WPs produce output for or consume input from other WPs, which also dictates order. Order of processing can be present without any exchange as well, e.g., an analysis requiring an intact target system state. Depending on the project, there can be WPs taking multiple inputs or generating multiple outputs. Even a uni-input/output interaction amongst the WPs, however, is sufficient to convey this paper's message.

Accordingly, the holistic overview of interactions between WPs can be represented with a \emph{WP dependency diagram}. Such a structure corresponds to common software project structures. Whichever terminology is considered, similar divisions and interactions are the case, e.g., packages, features, processes, components, issues, etc.

\subsubsection{Example: A structured multi-stakeholder research project}
\label{sec:example}
A representative case is the structure adopted in the ZORRO project, which focuses on data- and knowledge-driven diagnostics for industrial CPS~\cite{NWO:2023:ZORRO}. ZORRO is a typical research project focusing on algorithmic and software-based solutions. It involves academic, industrial, and applied research partners. The project is organised into seven work packages with the involvement of different users and stakeholders per work package\footnote{\url{https://zorro-project.nl/work-packages/}}:
\begin{itemize}
	\item WP~1 - Monitoring systems: Developing multi-level monitoring architectures, including meta-monitoring and embedded intelligence for system feedback.
	\item WP~2 - Knowledge representation: Creating formalised representations to support the sharing and reuse of diagnostic knowledge.
	\item WP~3 - Diagnostic algorithms: Designing advanced algorithms tailored for real-world CPS conditions, with an emphasis on reducing unplanned downtimes.
	\item WP~4 - Model-based systems engineering: Constructing modelling languages and frameworks to integrate diagnostics into engineering workflows.
	\item WP~5 - Demonstrators: Responsible for developing integrated demonstrators, ensuring the research components come together in practical showcases.
	\item WP~6 - Dissemination and adoption: In charge of outreach, training, and promoting long-term uptake of project results through guidelines and roadmaps.
	\item WP~7 - Management: Overseeing coordination, strategy, and compliance throughout the project lifecycle.
\end{itemize}

\begin{figure}[htbp]
	\centering
	\includegraphics[width=0.8\linewidth]{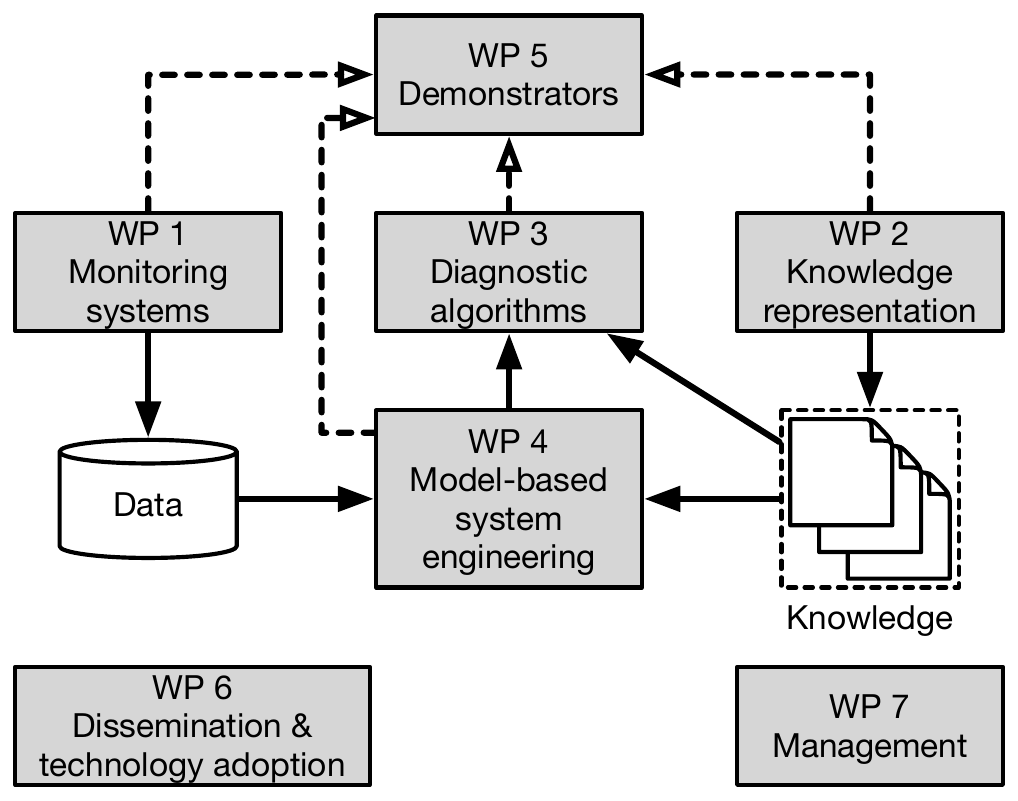}
	\caption{WP dependency diagram for the ZORRO project, covering direct dependencies with solid and uncertain/undefined demonstration WP input with dashed arrows.}
	\label{fig:wp_diagram_zorro}
\end{figure}

This structure exemplifies how different WPs contribute to a project's overall goals from distinct perspectives. For example, WP~5 depends on the technical maturity of outputs from WP~1 through WP~4. By extension, the technical maturity of these 4 WPs relies on interaction amongst them, e.g., system knowledge represented by WP~2 is to be taken advantage of by WP~3. In this particular case and with regards to the demonstrator specifics, potential misalignment in expectations, such as integration readiness or scope, could delay or jeopardise WP~5's outcomes. The WP dependency diagram for ZORRO is given in \Cref{fig:wp_diagram_zorro}. Note that the dependencies are only required if a holistic outcome is to be achieved, i.e., WPs can show independent functionality, e.g., using test data.

The ZORRO structure illustrates the need for early agreement on demonstrator types and target maturity levels, e.g., via a framework extending or complementing TRLs. Clear interfaces between WPs, well-defined data exchange protocols, and shared understanding of demonstration scope(s) are essential for smooth cross-WP collaboration.

\subsection{Targeted audience of a demonstrator}
Considering the aforementioned diversity of stakeholders involved in collaborative research projects, demonstrators rarely serve a single, homogeneous audience. Each stakeholder will evaluate a given demonstrator from multiple perspectives, i.e., maturity, scope, and usefulness. As a result, the notion of a \emph{successful demonstrator} is inherently stakeholder-dependent.

For instance, industrial partners prioritise clear applicability to real-world or production setups, which translates to an emphasis on robustness, scalability, system integration, and compliance with \emph{operational constraints}. In contrast, academic stakeholders may focus on conceptual novelty, methodological soundness, or reusability of the demonstrator, ideally to incorporate it in future research. Funding and oversight bodies on the other hand, tend to adopt a more balanced view, searching for both scientific contribution and practical applicability.

Given the rather differing expectations, a single demonstrator can seldom satisfy all involved parties. A demonstrator optimised for either operational realism and industrial validation, or experimental freedom and exploratory research, will sacrifice opposing aspects. As such, research projects ideally should produce multiple demonstrators, or variations of a single demonstrator with explicit scopes.

Therefore, clarifying the intended audience and by extension the purpose of a demonstrator early on, is essential in research projects. Such an identification helps align expectations and results in the right set of requirements. The variability in audience expectations directly impacts how demonstrator requirements are specified and prioritised, as discussed in the following section.

This variability also manifests at different levels of abstraction in requirements formulation. On the one hand, requirements must define the intended scope and ambition of demonstrators at the project-level. On the other hand, they must capture the detailed behaviour and constraints of individual demonstrators once their design is established. Distinguishing between these levels is essential to avoid ambiguity in both planning and implementation.

\subsection{Types of requirements}
Software requirements can be broadly divided into \emph{functional} and \emph{extra-functional} categories.

\paragraph*{Functional requirements}
Functional requirements define what the system should do, specifying the behaviour, features, and interactions that enable it to perform its intended tasks. These definitions can be in varying granularities. Common examples are processing data, executing Input/Output (IO) transactions, or providing user interfaces.

\paragraph*{Extra-functional requirements}
In contrast, extra-functional requirements, a.k.a., non-functional requirements\footnote{Terminology varies across the literature and can be context-dependent. Here, the term \enquote{extra-functional} is used to describe system qualities extending core functionality.}, define how the system should perform~\cite{Hochmuller:1999:TPIE}. These requirements cover qualities such as performance metrics, reliability levels, security, maintainability, and so forth. In short, functional requirements focus on system capabilities, while extra-functional requirements define operational constraints and standards to be adhered to by the system. Software development processes intended for production-ready systems involve both types for a successful outcome.

Following the two-layered view of requirements considered in this work, i.e., project-level demonstrator formulation and granular implementation requirements, the first layer is primarily characterised by whether functional and extra-functional requirements are considered in defining the intended demonstrator.

\section{Demonstrator taxonomy}
\label{sec:demonstrator_taxonomy}
We define a \emph{demonstrator taxonomy} for software-intensive research projects, corresponding to the project-level formulation of demonstrators as introduced in \Cref{sec:introduction}. Inspired by the Technology Readiness Level (TRL) scale described in \Cref{sec:background}, the taxonomy targets the first layer of requirements, i.e., the definition of demonstrator scope and ambition prior to implementation-specific considerations. While the TRL scale provides a well-established abstraction for assessing technology maturity, it lacks the granularity needed to capture the practical scope and integration realities of research projects with multiple Work Packages (WPs). By focusing on the context of software-intensive systems and industrial CPS, we provide a more concise taxonomy, relatable by researchers working on such projects. The taxonomy introduces five levels of demonstration, each defined in terms of:
\begin{itemize}
	\item Coverage scope: Which parts of the project are incorporated in the demonstrator, single WP, subset of WPs, or all WPs.
	\item Integration scope: Whether and how WPs interoperate in the demonstrator.
	\item Qualities covered: Whether only functional requirements are demonstrated or extra-functional requirements are also covered, e.g., successfully performing a task or performing the task while adhering to expected performance metrics (latency, throughput, \ldots).
\end{itemize}

This taxonomy is particularly relevant for software-intensive systems and focuses on industrial CPS research within Computer Science, where a significant portion of the effort lies in software integration, quality assurance, and stakeholder-driven requirements. Elaborations and a table view follows (\Cref{tab:demonstration_levels}).

We note that project proposals may explicitly state an intended demonstrator TRL level, e.g., TRL level~4--5 for a dissemination demonstrator, intended for public use. Our taxonomy, alongside the requirements elaboration framework, provides a structured way to assess whether such targets are feasible at the project-level and to clarify the corresponding requirements for achieving them. The taxonomy thus provides a structured abstraction for reasoning about demonstrators at the project-level, independent of implementation-specific design choices, which are addressed at the second layer of requirements.

\subsection{Work package perspectives}
To arrive at a complete taxonomy, both intra- and inter-WP perspectives are to be considered, i.e., functionality within a particular WP, independent of other WPs, and the functionality achieved through interaction.

\subsubsection{Intra-WP perspective}
From an intra-WP perspective, no apparent structure can be discovered with acceptable generality. This is due to the fact that individual WP descriptions and requirements are rather ad hoc (as it is expected to be). What we can consider though is a worst-case example, WP complexity-wise.

While most WPs, especially for engineering domains, collaborate with one industrial/external stakeholder, it is not uncommon to interact with multiple stakeholders. In such set-ups, it could well be the case that the requirements, experimentation, or even assigned workforce are disjoint and work on independent paths. Simply put, such an arrangement will render one WP into two or more (in extreme cases) sub-WPs.

\subsubsection{Inter-WP perspective}
The inter-WP perspective on the other hand is well-understood and consistently holds for most project set-ups. It is clear that each WP is expected to deliver functional and/or extra-functional capabilities. It is also assumed that most WPs are to interact with other WPs, usually following a straightforward producer-consumer relationship. There can be WPs that are \emph{islands}, interaction-wise, and WPs that consume from or produce for multiple corresponding WPs.

\subsection{Coverage levels in the taxonomy}
Accordingly, we distinguish 5 levels of demonstrator coverage, forming a hierarchy spanning from minimal proof-of-concept implementations to fully optimised, project-wide integrations.

\begin{figure}[htbp]
	\centering
	\includegraphics[width=0.9\linewidth]{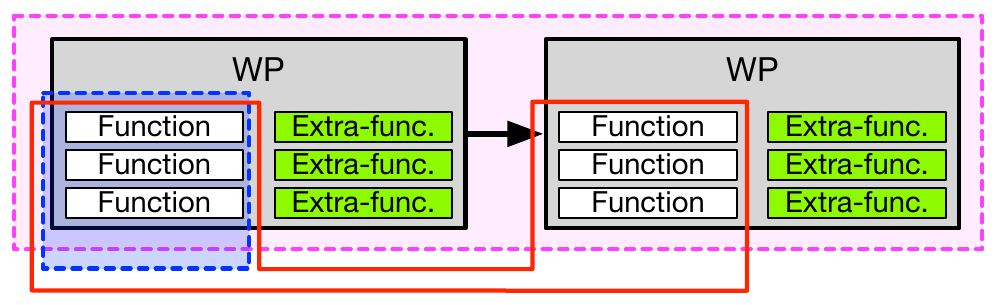}
	\caption{Visualising the coverage scope per different demonstrator level descriptions: Functional and extra-functional requirements (white and green blocks); Level~1 proof of concept (blue box); Level~2 proof of integration (red area); Level~3 optimised proof of integration (magenta box).}
	\label{fig:proof_levels}
\end{figure}

\paragraph*{1. Proof of concept}
It is the most common proof implementation, intended to showcase realisation of functional qualities. Given the considered project structure, it is strictly limited to a single WP's functional capabilities, as depicted with the blue box in \Cref{fig:proof_levels}.

Considering a level as \enquote{Optimised proof of concept} is also conceivable. However, since many extra-functional qualities are directly or indirectly related to interactions between WPs (modules in a software-centric design), we have opted to skip such a level.

\paragraph*{2. Proof of integration}
With a wider scope, the aim is to showcase realisation of functional qualities for a subset of WPs (2 or more), while \emph{interacting}. It can be seen as the red area in \Cref{fig:proof_levels}.

\paragraph*{3. Optimised proof of integration}
This level goes beyond the realisation of functional qualities and covers \emph{extra-functional} qualities as well. Examples are: performance, latency, throughput, explainability, resource efficiency, etc. The scope will still be limited to a subset of WPs, shown as the magenta box in \Cref{fig:proof_levels}.

\begin{figure}[htbp]
	\centering
	\includegraphics[width=\linewidth]{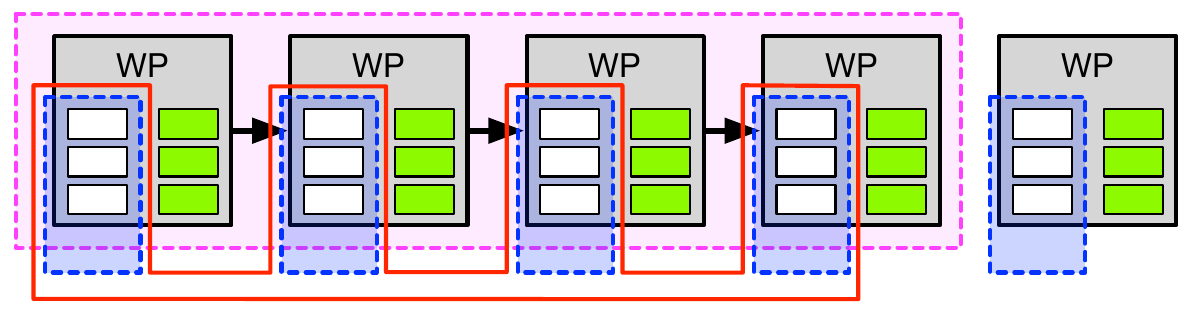}
	\caption{Visualising the project-wide view, covering all interacting WPs and in this case, an island WP: Functional and extra-functional requirements within any particular WP (white and green blocks); Level~1 proof of concept (blue boxes); Level~4 grand proof of integration (red area); Level~5 optimised grand proof of integration (magenta box).}
	\label{fig:grand_proof_levels}
\end{figure}

\paragraph*{4. Grand proof of integration}
A more comprehensive proof, similar to the proof of integration and covering the entirety of a project. Interaction between WPs is expected, excluding island WPs which may be present, as depicted with the red area in \Cref{fig:grand_proof_levels}.

\paragraph*{5. Optimised grand proof of integration}
A proof similar to level 3, but covering the entirety of a project, again, excluding island WPs. The scope for this level is visible as the magenta box in \Cref{fig:grand_proof_levels}.

Both demonstration levels~4 and~5 can be considered within higher TRL levels and will require extensive effort to realise. Such demonstrators could go beyond the expectations in a research project. A summary of the demonstrator taxonomy levels and their associated risk levels depending on the type of use-case involved is given in \Cref{tab:demonstration_levels}.
\begin{table*}[htbp]
    \centering
    \caption{Proposed levels of demonstration in research projects, each defined based on the covered project scope and covered qualities. Associated risks depending on the type of use-cases are listed per demonstration level.}
    \label{tab:demonstration_levels}
    \begin{tabularx}{\textwidth}{@{}cllllccc@{}}
        \toprule
        \multirow{2}{*}{\textbf{Level}} & 
    	\multirow{2}{*}{\textbf{Name}} & 
    	\multirow{2}{*}{\textbf{Coverage}} & 
    	\multirow{2}{*}{\textbf{Integration}} & 
    	\multirow{2}{*}{\textbf{Qualities Covered}} & 
        \multicolumn{3}{c}{\textbf{Use-case type risk association}} \\
        \cmidrule(l){6-8}
        & & & & & \textbf{Unified} & \textbf{Coordinated} & \textbf{Disparate} \\
        \midrule
        1   & Proof of concept            & Single WP      & No   & Functional					& Normal   & Normal    	& Normal \\
        2   & Proof of Integration (PoI)	& Subset of WPs  & Yes  & Functional               	& Moderate & Moderate  	& Impractical \\
        3   & Optimised PoI            	& Subset of WPs  & Yes  & Functional, Extra-func.		& High     & High      	& Impractical \\
        4   & Grand PoI             		& All WPs        & Yes  & Functional                	& Moderate & Impractical	& Impractical \\
        5   & Optimised grand PoI   		& All WPs        & Yes  & Functional, Extra-func.		& High     & Impractical	& Impractical \\
        \bottomrule
    \end{tabularx}
\end{table*}

\subsection{Use-case perspective}
In collaborative research projects, it is common for each industrial partner to contribute with its own use-case, i.e., a company-specific problem, system or scenario to which the demonstrator can be applied. Such a use-case is tailored to specific industrial or research interests of the partner. It is often the case that these differ significantly in scope, objectives, and technical requirements, and may not naturally align with those of other partners. As such, integration within the demonstrator can take several forms.

The use-case perspective does not alter the definitions of the demonstrator levels given in \Cref{tab:demonstration_levels}, but it can strongly influence the feasibility of achieving a given level in practice, by adding a new dimension: \emph{artefact readiness}, i.e., the readiness of the code/data associated with the use-case, as delivered by the industrial partner for integration in the demonstrator. Identifying use-case alignment early in the project lifecycle helps set realistic integration targets and guides requirements definition.

\subsubsection{Unified use-case}
All WPs contribute to a single, coherent scenario. Higher levels of demonstration, levels~4 and~5, are more easily achievable in this case. For this type, all levels of demonstration from \Cref{tab:demonstration_levels} are directly valid and feasible.

While having a single use-case facilitates the integration between WPs, it is highly unlikely to provide artefact readiness for all WPs. Received code/data might satisfy expectations of 1 or a subset of WPs at best, pointing to moderate risks involved. The artefact readiness is even a bigger challenge for levels covering extra-functional qualities. For these demonstration levels, additional information in the form of baselines or benchmarking results are expected, hence the associated risk is high.

\subsubsection{Coordinated multi-use-case}
Distinct use-cases share a common technological framework or interfaces, enabling meaningful integration despite differing end goals. For this type, all levels of demonstration from \Cref{tab:demonstration_levels} are adoptable and valid.

The same challenges as the Unified use-case type can be repeated here, however for a subset of WPs. As such, achieving demonstration levels~2 and~3 carry similar risks as the Unified Type. Optimisations covering extra-functional qualities will still require extra effort and higher artefact readiness. Levels~4 and~5 are considered not achievable by definition, since Coordinated multi-use-cases are not intended to cover integration of \emph{all} WPs within the project.

\subsubsection{Disparate use-cases}
Integration is primarily at the technology or component level, with demonstrators potentially organised as parallel showcases rather than a single, unified system. For this type, only the individual per WP level is valid. Disparate use-cases do not necessarily apply to all WPs.

On the one hand, the per WP risk is the lowest for Disparate use-cases, as considerations allowing integration are not to be covered. Any provided use-case will certainly include artefacts relevant to that single use-case. The quality of these artefacts still matter, hence the Normal risk association. Note that any demonstration level requiring integration between WPs is not achievable through Disparate use-cases, as indicated in \Cref{tab:demonstration_levels}.

\subsubsection{Risk level interpretation}
The qualitative risk associations reported in \Cref{tab:demonstration_levels} reflect the expected coordination effort, integration complexity, and artefact readiness required to realise each demonstrator level for a given use-case type. \enquote{Normal} risk denotes the baseline level of uncertainty typically expected when developing a demonstrator within a limited scope and without substantial integration constraints. In other words, any demonstrator development will involve at least a \enquote{Normal} risk level. \enquote{Moderate} risk reflects the additional challenges introduced by interoperability requirements between two or more WPs, including dependencies between software artefacts, interfaces, and stakeholder contributions. \enquote{High} risk further incorporates the effort required to satisfy extra-functional qualities, such as latency, throughput, robustness, or explainability, which generally necessitate additional optimisation, benchmarking, and validation activities. Finally, \enquote{Impractical} denotes combinations that are structurally incompatible with the corresponding use-case type, particularly where project-wide or cross-WP integration is not feasible by definition.

\subsection{Comparison to TRL scale}
\label{subsec:comparison_to_trl_scale}
Looking at \Cref{tab:trl_levels,tab:demonstration_levels}, there are two main shortcomings with the TRL scale:

First, the TRL scale does not clearly define the coverage scope, which is the WP dimension in research projects, while it does differentiate between stand-alone and integrated demonstrations. We clarify this dimension in 3 steps, i.e., single WP, subset of WPs, and the full set of WPs.

Second, the TRL scale works well for a single use-case, whereas research projects often involve multiple industrial partners and the use-case aspect forms a separate dimension. The aforementioned use-case types dictate limitations, which we have considered.

\section{Requirements elaboration framework}
\label{sec:framework}
Incorporating what has been explained so far, we design and present a structured framework for refining demonstrator requirements in the context of collaborative, software-intensive research projects. The framework is intended as an \emph{a priori} analysis tool to assess the feasibility of demonstrator objectives stated in proposals and to translate them into achievable, well-scoped requirements at the project-level. While the focus is on the first layer of requirements, i.e., demonstrator formulation, the framework also provides a basis for deriving more granular implementation requirements for individual demonstrators.

The approach integrates four key input dimensions: (i) TRL expectations per WP and for the demonstrator (from the project proposal); (ii) estimated achievable TRLs based on existing assets and planned work (from the WP planning); (iii) quality of artefacts, i.e., code and data (provided by collaborators/industrial partners); and (iv) the dependency structure between WPs (from the WP planning). By combining these inputs, the framework identifies realistic demonstration levels, highlights bottlenecks, and derives a consistent set of demonstrator requirements aligned with the selected level.

\begin{figure}[htbp]
    \centering
	\includegraphics[width=\linewidth]{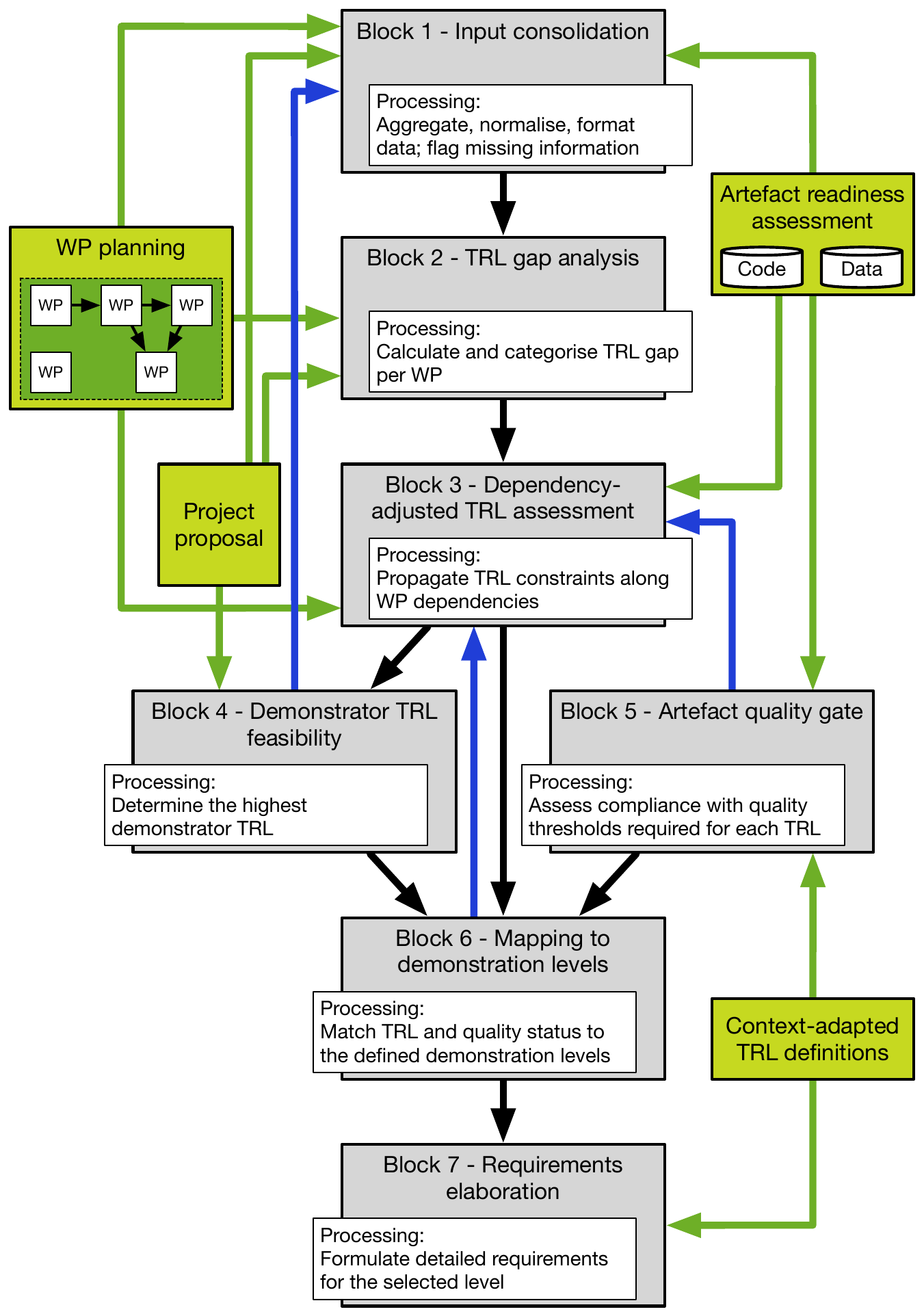}
    \caption{The demonstrator requirements elaboration framework is depicted. The diagram covers 7 steps with relevant processing, sources, as well as relations amongst these. Blue arrows are feedbacks to previous steps, green arrows are input from sources, and black arrows represent the sequential progression of blocks.}
    \label{fig:framework_diagram}
\end{figure}

Accordingly, the three key source documents are \emph{the project proposal}, \emph{the WP planning document}, and \emph{the code/data provisions from industrial partners}. Note that not every project is provided with a separate WP planning document. It is often time integrated with the project proposal itself. The framework consists of seven sequential analysis blocks. Each block is defined in terms of its inputs, processing activities, outputs, and any feedback relationships to preceding blocks. \Cref{fig:framework_diagram} illustrates the overall flow.
A detailed overview of the analysis blocks, covering input, processing action, output, and possible feedback to previous blocks is listed in \Cref{tab:block_descriptions}.
\begin{table*}[htbp]
    \centering
    \caption{Overview analysis blocks from the demonstrator requirements elaboration framework, describing inputs, processing, outputs, and feedback per block.}
    \label{tab:block_descriptions}
    \begin{tabularx}{\textwidth}{@{}p{3.0cm}X@{}}
        \toprule
        \textbf{Block} & 
        \textbf{Description} \\
        \midrule
        1 – \newline
        Input consolidation & 
        \textbf{Input:} TRL target per WP (project proposal); TRL target for the demonstrator or demonstrators (project proposal); estimated achievable TRL per WP at project completion (WP planning); artefact readiness assessments per WP, based on code/data (industrial partner submission); WP dependency diagram (WP planning). \newline
        \textbf{Processing:} Aggregate and normalise input data into a consistent format; flag missing or uncertain information. \newline
        \textbf{Output:} Complete input dataset; list of missing or uncertain elements. \newline
        \textbf{Feedback:} None. \\
        \midrule
        2 – \newline
        TRL gap analysis & 
        \textbf{Input:} TRL targets (project proposal); estimated achievable TRL per WP (WP planning). \newline
        \textbf{Processing:} For each WP, calculate the TRL gap between TRL targets and estimated achievable TRLs; categorise as on track, minor gap, or major gap. \newline
        \textbf{Output:} Gap analysis table for all WPs. \newline
        \textbf{Feedback:} None. \\
        \midrule
        3 – \newline
        Dependency-adjusted TRL assessment & 
        \textbf{Input:} WP dependency diagram (WP planning); TRL gap analysis (Block~2 output); code/data readiness assessments (industrial partner submission). \newline
        \textbf{Processing:} Propagate TRL constraints along WP dependencies; adjust achievable TRL per WP based on upstream limitations and artefact quality; identify bottlenecks and isolated WPs. \newline
        \textbf{Output:} Adjusted TRL map; list of critical and bottleneck WPs. \newline
        \textbf{Feedback:} None. \\
        \midrule
        4 – \newline
        Demonstrator TRL \newline
        feasibility & 
        \textbf{Input:} Adjusted TRL map (Block~3 output); demonstrator TRL target (project proposal). \newline
        \textbf{Processing:} Determine the highest demonstrator TRL achievable given WP constraints; compare with target; record shortfalls and possible mitigations. \newline
        \textbf{Output:} Achievable demonstrator TRL; TRL shortfall analysis. \newline
        \textbf{Feedback:} If TRL information is incomplete, return to Block~1 or Block~2. \\
        \midrule
        5 – \newline
        Artefact quality gate & 
        \textbf{Input:} Code/data readiness assessments (industrial partner submission); TRL definitions for software-intensive systems (context-adapted reference). \newline
        \textbf{Processing:} Assess compliance with quality thresholds required for each TRL stage; identify risks to integration and non-functional testing. \newline
        \emph{Note: The assessment is performed by the artefact users, i.e., the research team utilising any particular artefact.} \newline
        \textbf{Output:} Quality compliance report; recommended corrective actions. \newline
        \textbf{Feedback:} If quality gaps affect dependencies, return to Block~3. \\
        \midrule
        6 – \newline
        Mapping to \newline
        demonstration levels & 
        \textbf{Input:} Achievable demonstrator TRL (Block~4 output); adjusted TRLs per WP (Block~3 output); artefact quality compliance report (Block~5 output). \newline
        \textbf{Processing:} Match TRL and artefact quality status to the defined demonstration levels (\Cref{sec:demonstrator_taxonomy}); select the highest level attainable without major additional work. \newline
        \textbf{Output:} Recommended demonstration level; justification; constraints report. \newline
        \textbf{Feedback:} If aiming for a higher level, return to Block~3 to explore TRL upgrades. \\
        \midrule
        7 – \newline
        Requirements \newline
        elaboration & 
        \textbf{Input:} Recommended demonstration level (Block~6 output); TRL-specific requirements (context-adapted reference); demonstration level definitions (\Cref{sec:demonstrator_taxonomy}). \newline
        \textbf{Processing:} Formulate demonstrator requirements consistent with the selected level, covering functional, extra-functional, integration, and validation aspects at the project-level; outline implications and influences for subsequent implementation-specific requirement refinement, which will be carried out during the research. \newline
        \textbf{Output:} Comprehensive demonstrator requirements specification; WP improvement plan. \newline
        \textbf{Feedback:} None. \\
        \bottomrule
    \end{tabularx}
\end{table*}

The framework thus operationalises the transition from high-level demonstrator ambitions to structured requirements at the project-level, while maintaining a clear separation from implementation-specific design decisions.

\subsection{Framework artefacts and usage}
While \Cref{tab:block_descriptions} lists a detailed step-wise description of the framework, providing a separate view of the method in terms of consumed and produced artefacts is advantageous. This perspective enables practical adoption and reuse of the framework across projects.

\paragraph*{Inputs}
The framework operates on three primary sources: (i) the project proposal, providing TRL targets and demonstrator ambitions; (ii) the WP planning document, capturing estimated achievable TRLs and dependency structures; and (iii) artefact provisions from industrial partners, i.e., code and data used for validation and integration. These inputs are consolidated into a unified dataset in Block~1.

\paragraph*{Intermediate artefacts}
During execution, the framework produces a set of intermediate artefacts that guide the analysis:
\begin{itemize}
    \item TRL gap analysis tables, highlighting mismatches between targets and expected outcomes.
    \item Dependency-adjusted TRL maps, capturing propagated constraints across WPs.
    \item Artefact quality reports, assessing readiness of code/data for integration and validation.
    \item Bottleneck and constraint lists, identifying critical limitations affecting demonstrator feasibility.
\end{itemize}

\paragraph*{Outputs}
The primary outputs of the framework are:
\begin{itemize}
    \item A recommended demonstrator level, aligned with achievable TRLs and integration constraints.
    \item A structured set of demonstrator requirements at the project-level, covering functional, extra-functional, integration, and validation aspects.
    \item A WP-level improvement plan, outlining actions required to achieve higher demonstrator levels, if desired.
\end{itemize}

\paragraph*{Templates and representations}
The artefacts produced by the framework can be represented using simple and reusable templates, such as tabular TRL mappings per WP, dependency diagrams, and requirement specification tables structured by demonstrator level. These representations are intentionally lightweight and abstract, allowing adaptation to different project formats without imposing strict tooling requirements.

\subsection{Artefact readiness assessment}
To improve consistency and reduce subjective interpretation in Block~5 of the framework (\emph{Artefact quality gate}), a lightweight artefact readiness assessment can be considered. The assessment serves as a common baseline for evaluating the readiness of code and data artefacts. It is intentionally lightweight and does not aim to replace formal software quality assurance or certification procedures. Instead, it provides a practical mechanism for estimating integration risks, validation feasibility, and the expected effort required to achieve higher demonstrator levels.

The assessment considers three complementary dimensions: implementation quality, data quality, and integration readiness. These dimensions reflect common challenges encountered in collaborative software-intensive research projects, particularly where externally provided artefacts are reused across WPs. A simple readiness classification can be summarised as follows:
\begin{itemize}
    \item Low readiness: Prototype or incomplete artefacts with limited documentation, unstable interfaces, incomplete datasets, or substantial manual integration effort required.
    \item Medium readiness: Functional and partially documented artefacts with representative datasets, reasonably stable interfaces, and limited reproducibility or automation support.
    \item High readiness: Validated and reproducible artefacts with clear documentation, curated datasets, stable interfaces, and well-defined integration or deployment procedures.
\end{itemize}

The readiness assessment is intended to support feasibility analysis rather than strict acceptance criteria. Artefacts assessed at lower readiness levels may still be usable within limited-scope demonstrators, but they are expected to introduce increased integration and validation risks. In some cases, lower artefact readiness may be compensated through additional support from the artefact providers, e.g., industrial partner teams.

\section{Discussion and example projects}
\label{sec:discussion}
We cover pros and cons, as well as different practical details by applying our framework to research projects ZORRO and PrimaVera.

\subsection{The case of ZORRO}
\label{subsec:the_case_of_zorro}
We have introduced the structure for the ZORRO project in \Cref{sec:example}, covering the projects WP descriptions and WP dependency diagram (\Cref{fig:wp_diagram_zorro}). With this project, there is no separate WP planning included and all available information is provided within the project proposal.

\subsubsection{ZORRO WP dependency walk-through}
To elaborate the potential interaction that different WPs might have, we describe the diagram from \Cref{fig:wp_diagram_zorro} in a concrete scenario. The presented scenario can be considered for any industrial CPS from the project's company use-case portfolio, e.g., an ITEC die bonder machine, or a Canon production printer:
\begin{itemize}
	\item WP~1's involvement: Monitoring subsystems collect and consolidate data representing machine behaviour, covering numerous metrics, e.g., voltage, current, torque, temperature, etc., and time. The consolidation may include synthetic and semi-synthetic data generation, potentially data fusion, as well as sensor behaviour studies. For instance sensor noise simulation can be covered.
	\item WP~2's involvement: The knowledge of the system at hand is extracted and consolidated in suitable formats, e.g., a knowledge graph, from machine-specific artefacts, or expert input.
	\item WP~4's involvement: Having the domain knowledge in general and the machine's internal operations in particular, model-based data sanitisation, augmentation, imputation, etc., is applied. Consolidated knowledge from WP~2 can be considered as an input (\Cref{fig:wp_diagram_zorro}), for instance describing correct data formats and hierarchical relations between different data structures, or declared associations of data segments with machine operations are to be taken into account.
	\item WP~3's involvement: Algorithms, data processing pipelines, and ML models perform Fault Detection and Isolation (FDI), consuming data passed on from WP~4. Consolidated knowledge from WP~2 is to be used in both the actual design of algorithms and from a high-level perspective, to decide on high-value faults and anomalies.
\end{itemize}

\subsubsection{Considered demonstrators}
The proposal document defines the purpose of demonstrators as tools to \enquote{incorporate scientific findings for dissemination to various audiences}. A blanket target of TRL levels 4--5 is set early on. Further into the document, two categories of demonstrators have been defined: industrial demonstrators and dissemination demonstrators. The dissemination demonstrators are defined as \enquote{proof-of-concept system}, while the industrial demonstrator is denoted as \enquote{based on industrial use-cases}. For the latter, two separate demonstrator instances are targeted. Additionally, it is clearly indicated that the dissemination demonstrator shall focus on increasing the TRL level for work package outputs.

\begin{figure}[htbp]
	\centering
	\includegraphics[width=\linewidth]{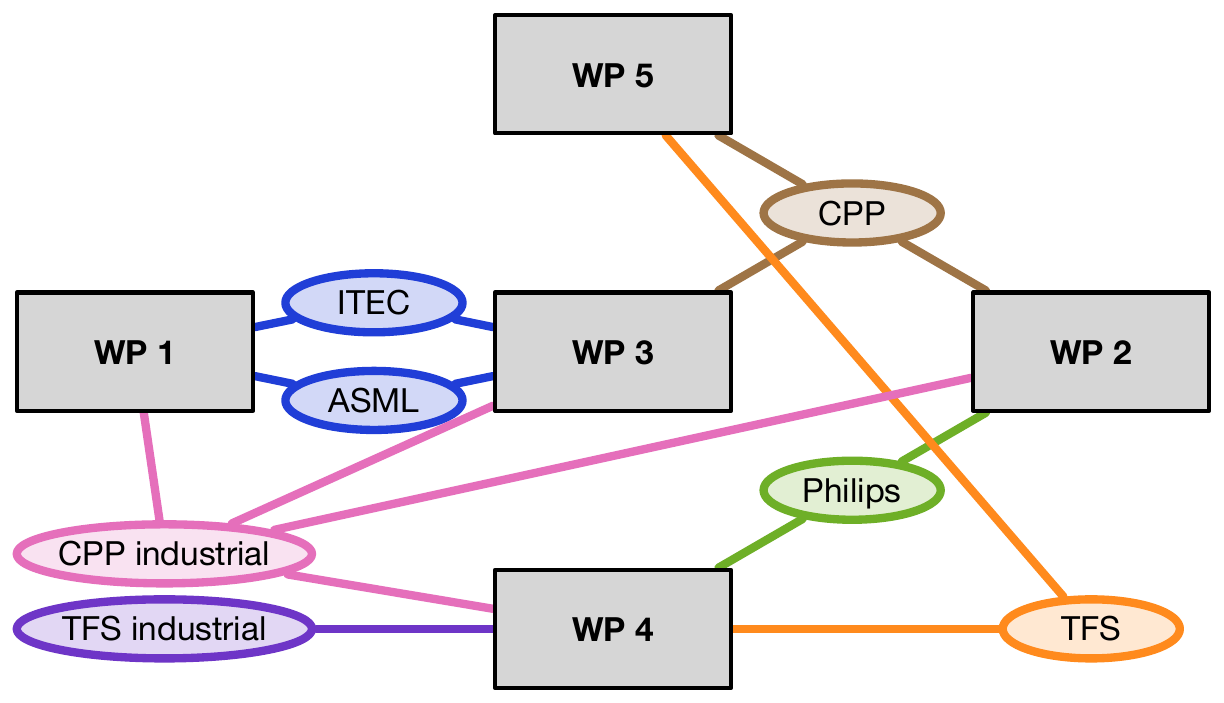}
	\caption{Use-case associations at both WP and intended industrial demonstrator levels for the ZORRO project. Company use-cases with identical WP association are drawn with the same colour.}
	\label{fig:wp_company_zorro}
\end{figure}

More granular descriptions are provided in the WP planning, which is often appended to the project proposal itself. Specifically, the planning for WP~5 (in charge of demonstrators) reiterates the target TRL scale as 4--5. The dissemination demonstrator is described as a \enquote{simulated production environment}, covering all WPs, \emph{implying integration} and thus TRL level 5. The industrial demonstrator is to apply partner use-cases, i.e., industrial partner artefact utilisation, CPP for WPs~1--4 and TFS for WP~4. Full coverage of modules using industrial artefacts points to TRL level~6. Use-case associations within the ZORRO project plus a legend of the involved companies are given in \Cref{tab:wp_company_zorro}.
\begin{table}[htbp]
    \centering
    \caption{Industrial use-case associations in ZORRO.}
    \label{tab:wp_company_zorro}
    \begin{tabularx}{\linewidth}{@{}cll@{}}
        \toprule
        \textbf{WP} & 
        \textbf{Collaboration use-case(s)} &
        \textbf{Industrial demo.} \\
        \midrule
        1 	& ASML, ITEC 		& CPP \\
        2 	& CPP, Philips 		& CPP \\
        3 	& ASML, CPP, ITEC 	& CPP \\
        4 	& Philips, TFS 		& CPP, TFS \\
        5 	& CPP, TFS 			& n/a \\
        \bottomrule
    \end{tabularx}
    \vspace{0.25em}
    \begin{minipage}{\linewidth}
        \footnotesize
        \raggedright
        \textbf{Legend.} \\
        \hspace{1.5em}ASML: ASML Holding (\url{https://www.asml.com/en}); \\
        \hspace{1.5em}CPP: Canon Production Printing (\url{https://cpp.canon/}); \\
        \hspace{1.5em}ITEC: ITEC Equipment (\url{https://www.itecequipment.com/}); \\
        \hspace{1.5em}Philips: Philips Medical Systems (\url{https://www.philips.nl/healthcare}); \\
        \hspace{1.5em}TFS: Thermo Fisher Scientific (\url{https://www.thermofisher.com/nl/en/home.html}).
    \end{minipage}
\end{table}
A visual diagram illustrating use-case associations at both WP level and the intended (per project proposal) industrial use-case associations within the ZORRO project, is provided in \Cref{fig:wp_company_zorro}.

\subsubsection{Applying the framework}
Application of the proposed framework (\Cref{fig:framework_diagram}) then yields the following insights:

\paragraph*{Block~1}
The above granular information from all the considered sources (\Cref{tab:block_descriptions}) is collected. Not every project provides all expected items, which can be compensated for through basic deduction. We have done this to clarify targeted TRL levels for different demonstrators listed in the project proposal and WP planning. Information around industrial partner artefact readiness is also not necessarily available in full. At this stage, we can only collect use-case associations.

\paragraph*{Block~2}
While no individual TRL targets are provided per WP, we can consider the blanket TRL~4--5 as an indicator. Considering our deduced TRL~5 for the dissemination demonstrator, we can interpret the expected maturity per WP to be around TRL~4. At this stage, this is a minor gap, as the difference would be the integration aspect. With regards to the industrial demonstrator, TRL~6, the gap is already major.

\paragraph*{Block~3}
When it comes to code/data readiness, availability is a major factor, as can be observed from the use-case associations listed in \Cref{tab:wp_company_zorro}. To achieve integration demonstrations, there needs to be an alignment between the WP dependency diagram and use-case associations. For instance, while WPs~1 and 3 are associated with common use-cases from ASML and ITEC, promising a coordinated multi-use-case type scenario, these WPs have no direct dependency relationship. To be able to demonstrate integration, WP~4 as the connecting WP between WPs~1 and 3 (\Cref{fig:wp_diagram_zorro}) has to be included, which is not associated with ASML or ITEC use-cases. This highlights a structural misalignment between dependency requirements for integration and the distribution of available use-case artefacts.

\paragraph*{Block~4}
Realising both dissemination and industrial demonstrators within ZORRO with targeted TRLs is highly challenging under current circumstances. The TRL scale does not explicitly accommodate partial integration across subsets of WPs, making it difficult to represent intermediate demonstrator states under the given constraints. As such, considering partial use-case associations, both TRLs~5 and 6 are out of reach. Even though there is a dedicated WP with a dedicated use-case, from CPP, the reliance on the results (software modules) from WPs~1--4 is an obstacle. Adjustments to WP planning have to be made.

\paragraph*{Block~5}
As mentioned in Block~1, exact code/data readiness is not clear at this stage in the project, except use-case associations, which we will rely on.

\paragraph*{Block~6}
Given the use-case associations, a mix of coordinated multi-use-case and disparate use-case types exist. By mapping the demonstrator targets to our defined levels of demonstration from \Cref{sec:demonstrator_taxonomy}, the taxonomy accommodates partial integrations explicitly, corresponding to level~2 (proof-of-integration) and potentially level~3 (optimised proof-of-integration). While the use-case from CPP initially promises the possibility of a grand proof-of-integration, the use-case is not associated with WPs~1 and 4. In essence, the target needs to be adjusted to level~2 (proof-of-integration) and possibly level~3 (optimised proof-of-integration).

\paragraph*{Block~7}
Following the identification of feasible demonstration levels, demonstrator requirements can be formulated at the project-level, covering functional, extra-functional, integration, and validation aspects consistent with the selected level. In this case, the primary implication is that requirements must be aligned with level~2 (proof-of-integration) or level~3 (optimised proof-of-integration) demonstrators, rather than the initially targeted higher TRLs. A detailed instantiation of such requirements is omitted, as it is highly case-specific and beyond the scope of this paper.

\paragraph*{Action}
Following the need for adjustments identified in Blocks~4 and 6, two main options are present. Either the WP planning has to be adjusted towards level~4 (grand proof-of-integration) by enforcing the association of CPP use-case in WPs~1 and 4; or alternatively, the targeted industrial demonstrator is to be redefined as multiple level~2 or level~3 demonstrations, each covering a subset of WPs. Needless to say, these demonstrators cannot all be based on the CPP use-case and should adopt the commonly associated use-case from respective WP subsets.

\subsubsection{Implications for proposal composition}
This case illustrates how the framework reveals misalignments between demonstrator ambitions, WP dependencies, and artefact availability, enabling early adjustment of demonstrator scope and expectations. While applied here retrospectively, the primary intent of the framework is \emph{to support demonstrator formulation at proposal composition time}. At that stage, research outcomes, implementation details, and even concrete artefacts are largely unknown, making it difficult to define realistic demonstrator targets. By structuring available information, assumptions, and expected dependencies, the framework supports the formulation of demonstrator requirements at the project-level, i.e., the first layer of requirements (as introduced in \Cref{sec:introduction}). The case study thus serves to showcase how such misalignments can be identified early on, rather than after project initiation, where corrective actions are significantly more costly.

\subsection{The case of PrimaVera}
\label{subsec:the_case_of_primavera}
As a second example and a project in its final stages, we consider the PrimaVera project, which addresses predictive maintenance across complex systems using data-driven and cross-level optimisation methods~\cite{NWO:2019:PrimaVera}.

\subsubsection{WP descriptions}
It is organised into 6 technical work packages plus 2 additional packages focused on dissemination and management\footnote{\url{https://primavera-project.com/work-packages/}}:
\begin{itemize}
	\item WP~1 – Data acquisition: Defining what to measure and how to gather data to optimise its quality for predictive maintenance; exploring novel data sources and establishing decision-support mechanisms.
	\item WP~2 – Data processing and diagnosis: Converting raw sensor data into actionable insights. Developing real-time automated methods for data validation, missing-value imputation, outlier detection, and isolating failure-related signals from operational condition changes.
	\item WP~3 – Prognostic algorithms: Creating scalable and accurate prognostic models to predict component and system failures. Generating key performance indicators like Remaining Useful Life (RUL), time to first failure, reliability, and availability, feeding into optimal decision-making in later stages.
	\item WP~4 – Maintenance and logistics optimisation: Designing robust, efficient methods for large-scale maintenance planning and service logistics. Integrating prognostic indicators from WP~3, combining maintenance and logistics planning, with adaptability and robustness at the core.
	\item WP~5 – Organisational behaviour and human decision making: Exploring user requirements and design principles for decision-support tools in predictive maintenance. Aiming to create user-centred maintenance systems integrated effectively into business processes, and new guidelines for implementation.
	\item WP~6 – Predictive maintenance demonstrators: Building three demonstrators that integrate findings from WP~1 to 5. These act as modular building blocks, showcasing the practical value of the research for industrial applications and facilitating exploitation and uptake.
	\item WP~7 – Dissemination and knowledge utilisation: Focusing on outreach, translating results into usable forms, and promoting broader uptake of project findings.
	\item WP~8 – Management: Ensuring effective coordination, strategic oversight, and operational management across the project lifecycle.
\end{itemize}

\begin{figure}[htbp]
	\centering
	\includegraphics[width=0.8\linewidth]{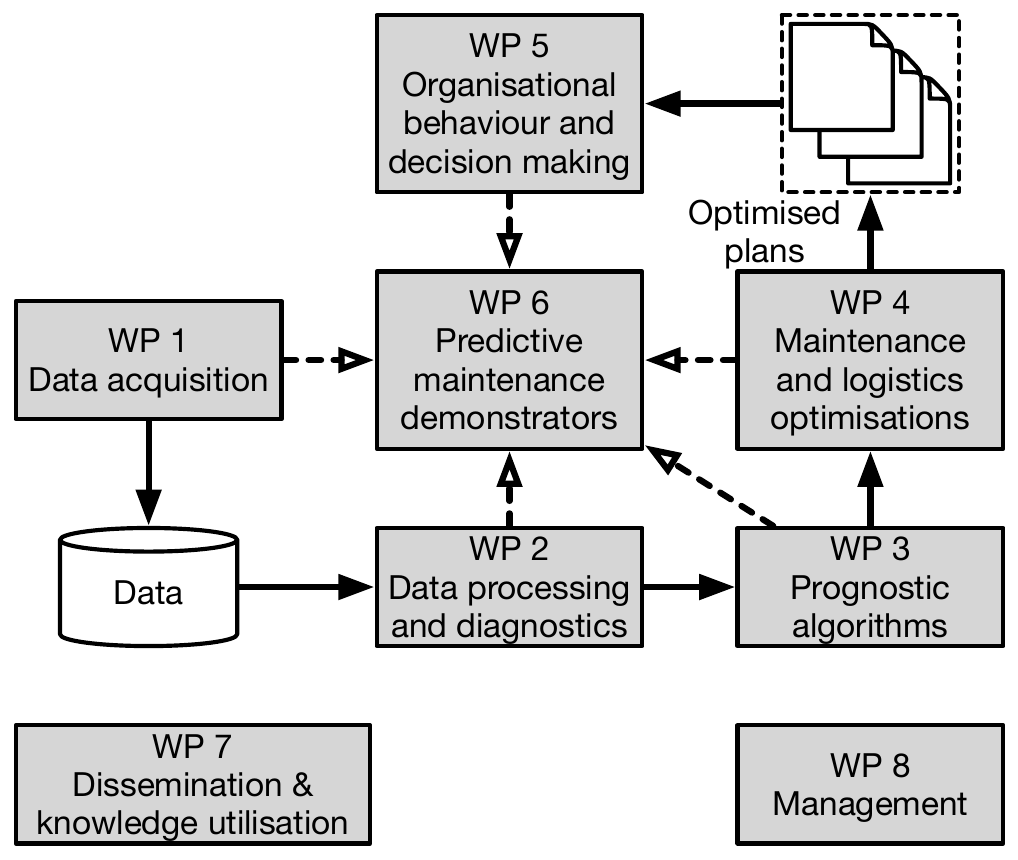}
	\caption{WP dependency diagram for the PrimaVera project, covering direct dependencies with solid arrows and input to the demonstration WP with dashed arrows.}
	\label{fig:wp_diagram_primavera}
\end{figure}

The WP dependency diagram for PrimaVera is given in \Cref{fig:wp_diagram_primavera}. A concrete scenario to elaborate and crawl the diagram is given next. Note that this scenario is hypothetical and only intended to provide an understanding of the interactions.

\subsubsection{PrimaVera WP dependency walk-through}
To elaborate the potential interaction that different WPs might have, we describe the diagram from \Cref{fig:wp_diagram_primavera} in a concrete scenario. The presented scenario is specific to the naval use-case from the project and brings different WPs together around a digital twin representation of a naval vessel, as a test bench:
\begin{itemize}
	\item WP~1's involvement: Data sensing algorithms, sensor network designs, data labelling strategies, and data conversion algorithms are incorporated to collect operational and maintenance data from component use-cases. Considered component data covers centrifugal pumps, bearings, vessel hull corrosion, vessel hull cracks, and impact effects.
	\item WP~2's involvement: To achieve diagnostics, several algorithms and methods are developed to detect faults, e.g., anomaly detection, or to assess the system health, e.g., impact and structural damage assessment.
	\item WP~3's involvement: Prognostic algorithms are built around three approaches: data-driven, model-based, and hybrid. Considered components cover centrifugal pumps, vessel hull corrosion, and vessel hull cracks.
	\item WP~4's involvement: Aspects such as maintenance optimisation and spare part inventory management are studied and applied to cases covering centrifugal pumps, bearings, and vessel hull corrosion.
	\item WP~5's involvement: User-friendly decision support algorithm design and organisational implementation strategies are covered and applied to centrifugal pumps, vessel hull corrosion, and vessel hull cracks.
\end{itemize}

\subsubsection{Considered demonstrators}
Similar to the ZORRO project, there is no separate WP planning included. The proposal document describes \enquote{three closely interlinked large-scale field-lab demonstrators}. The term \enquote{interlinked} indicates that even the different demonstrators are expected to implement some form of integration, which is not clarified. There are no indications of per-WP demonstrators, nor any TRL levels for any type of demonstration.

Demonstrator~1 is a health assessment and prognostics tool, which will implement the data collection, diagnostics and prognostic methods into a practical software tool. Demonstrator~2 is a planning and maintenance tool, which will implement the maintenance and supply chain optimisation into a user-friendly software tool. Demonstrator~3, the most comprehensive one, is a digital twin to support ship maintenance. The digital twin is a computer model representing the physical components and functions of (part of) a real ship, as well as its degradation behaviour. Though no TRL level is indicated, what we immediately notice from the jargon is the affinity towards tooling and targeted industrial maturity. The concept of a digital twin closely matches the \emph{high-fidelity simulated environment} and pointing to TRL~7, a NASA designation~\cite{Kimmel:2020:TRAB}.

\begin{figure}[htbp]
	\centering
	\includegraphics[width=\linewidth]{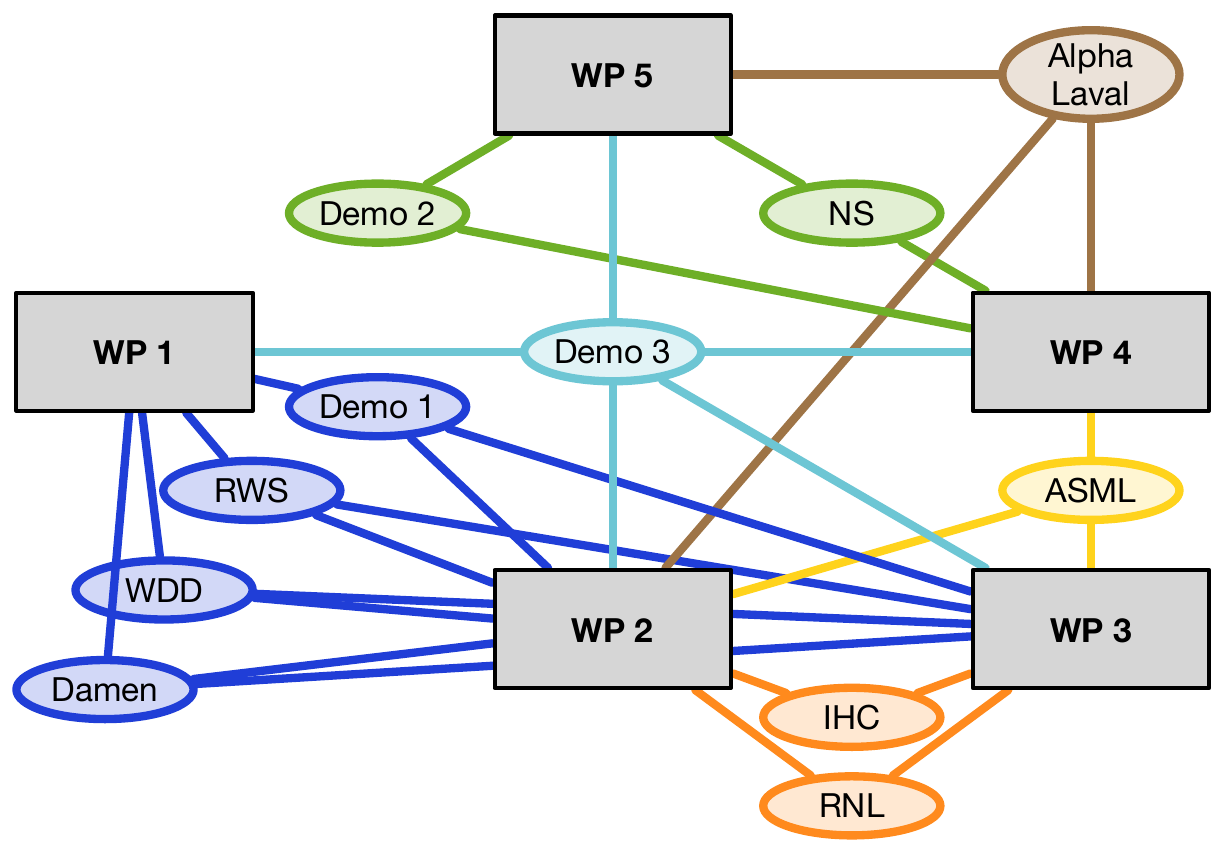}
	\caption{Use-case associations at both WP and intended industrial demonstrator levels for the PrimaVera project. Company use-cases with identical WP association are drawn with the same colour.}
	\label{fig:wp_company_primavera}
\end{figure}

Further descriptions under WP~6, responsible for demonstrators, provides targeted WP coverage per each demonstrator. These coverages, alongside use-case associations within the PrimaVera project are given in \Cref{tab:wp_company_primavera}. A legend of the involved companies is included as well. Not every single industrial partner is listed in the table. There are partners who are not clearly associated with any WP, as well as partners who do not contribute with code/data.
\begin{table}[htbp]
    \centering
    \caption{Industrial use-case associations in PrimaVera.}
    \label{tab:wp_company_primavera}
    \begin{tabularx}{\linewidth}{@{}cXl@{}}
        \toprule
        \textbf{WP} & 
        \textbf{Collaboration use-case(s)} &
        \textbf{Industrial demo.} \\
        \midrule
        1 	& RWS, WDD, Damen 									& Demo. 1, 3 \\
        2 	& RWS, WDD, ASML, Damen, IHC, Alfa Laval, RNL 	& Demo. 1, 3 \\
        3 	& RWS, WDD, ASML, Damen, IHC, RNL 					& Demo. 1, 3 \\
        4 	& ASML, NS, Alfa Laval 							& Demo. 2, 3 \\
        5 	& NS, Alfa Laval  									& Demo. 2, 3 \\
        \bottomrule
    \end{tabularx}
    \vspace{0.25em}
    \begin{minipage}{\linewidth}
        \footnotesize
        \raggedright
        \textbf{Legend.} \\
        \hspace{1.5em}Alfa Laval: Alfa Laval Benelux (\url{https://www.alfalaval.nl/nl/}); \\
        \hspace{1.5em}ASML: ASML Holding (\url{https://www.asml.com/en}); \\
        \hspace{1.5em}Damen: Damen Shipyards Group (\url{https://www.damen.com/}); \\
        \hspace{1.5em}IHC: Royal IHC (\url{https://www.royalihc.com/}); \\
        \hspace{1.5em}NS: Netherlands Railways (\url{https://www.ns.nl/en}); \\
        \hspace{1.5em}RNL: Royal Netherlands Navy (\url{https://english.defensie.nl/organisation/navy}); \\
        \hspace{1.5em}RWS: Rijkswaterstaat (\url{https://www.rijkswaterstaat.nl/en}); \\
        \hspace{1.5em}WDD: Water Board de Dommel (\url{https://www.dommel.nl/}).
    \end{minipage}
\end{table}
A visual diagram illustrating use-case associations at both WP level and the intended (per project proposal) industrial use-case associations within the PrimaVera project, is provided in \Cref{fig:wp_company_primavera}.

The proposal, at the time, did not include which use-cases were to be used for demonstrators. It turned out that Demo.~1 and Demo.~2 were decided to be based on cases and artefacts from RWS and NS, respectively. Demo.~3 points to RNL, Damen, IHC and Alfa Laval as the only industrial partners involved with naval use-cases. It is explicitly indicated that Demo.~3 \enquote{will integrate all methods developed in the WPs~1--5 into a digital twin of a ship}, which translates to a \emph{grand proof of integration} and a high TRL. Lastly, PrimaVera has resulted in a \enquote{product catalogue}\footnote{\url{https://primavera-project.com/catalogue/}}, i.e., a collection of per WP demonstrators. These correspond to level~1 in our taxonomy, proof of concept, falling short of the three targeted industrial Demos.

\subsubsection{Applying the framework}
Application of the proposed framework (\Cref{fig:framework_diagram}) then yields the following insights:

\paragraph*{Block~1}
The above granular information from all the considered sources is collected. Considering the language, we estimate TRL~6 as the target for demonstrators~1 and 2, and TRL~7 as the target for demonstrator~3. Information around industrial partner artefact readiness is not available at this stage, but use-case associations are collected.

\paragraph*{Block~2}
Given the high deduced TRL levels for the three comprehensive demonstrators, a gap is immediately recognisable when compared to the expected maturity of individual WPs in research-driven settings, where lower TRLs are the norm.

\paragraph*{Block~3}
For PrimaVera, a good alignment has been achieved between the industrial partners involved in Demo.~1 and 2 with WP associations (\Cref{tab:wp_company_primavera}). RWS is involved in WPs~1--3 and NS is involved in WPs~4 and 5. The same is not the case for Demo.~3 where none of the naval use-case providers are associated with all 5 WPs, pointing to an availability issue. Note that these alignments were not clarified within the proposal and are later ad hoc adjustments, e.g., RWS was chosen for Demo.~1 organically, precisely because of its involvement in WPs covered by Demo.~1. Block~3 of our framework would have detected and enforced adjustments \emph{early on}. The WP diagram for PrimaVera defines WP relationships sequentially, from 1 to 5, which is simpler than the case of ZORRO. This highlights that alignment between use-case artefacts and WP dependencies was achieved reactively, whereas the framework would have enabled such alignment to be enforced during proposal composition.

\paragraph*{Block~4}
The deduced (TRL scale is not considered in the proposal) TRL levels for Demo.~1 and 2 were TRL~6, followed by TRL~7 for Demo.~3. While it is not clarified, it can be assumed that per WP demonstrations will not reach anywhere close to such high TRL levels, since the bulk of the work is performed by research and academic entities. Relying on our deduced TRLs and considering that Demo.~3 is expected to \emph{integrate} outcomes of Demo.~1 and~2, the TRL mismatch is apparent, demanding adjustments to WP planning.

\paragraph*{Block~5}
Exact code/data readiness were not clarified at the time.

\paragraph*{Block~6}
Mapping to our proposed demonstration levels, Demo.~1 and 2 correspond to level~3 (optimised proof of integration), due to expected integration and the fact that a tool is to be produced, i.e., emphasis on optimised operation. For Demo.~3, a demonstration level of 5 (optimised grand proof-of-integration) can be designated, since a grand integration and high-fidelity digital-twin in the loop are expected. Clearly, these classifications highlight the ambitious nature of the targeted demonstrators.

\paragraph*{Block~7}
Following the identification of feasible demonstration levels, demonstrator requirements can be formulated at the project-level, covering functional, extra-functional, integration, and validation aspects consistent with the selected level. In this case, the primary implication is that requirements for Demo.~1 and 2 align with level~3 demonstrators, while Demo.~3 would require level~5 requirements, which introduces significant feasibility challenges. A detailed instantiation of such requirements is omitted, as it is highly case-specific and beyond the scope of this paper.

\subsubsection{Implications for proposal composition}
This case further illustrates the role of the framework as a design-time instrument for demonstrator formulation. In PrimaVera, several key decisions, such as the selection of use-case providers for demonstrators and the alignment between WP dependencies and industrial artefacts, were made after project initiation. While these adjustments led to workable demonstrator definitions, they reflect a reactive process.

The framework would have enabled a systematic and structured evaluation of demonstrator feasibility at proposal composition time, particularly by exposing mismatches between targeted TRLs, WP-level maturity, and use-case availability. This is especially relevant for highly ambitious demonstrators such as Demo.~3, where assumptions regarding integration and artefact readiness play a critical role. As such, the case showcases how the framework supports the formulation of demonstrator requirements at the project-level under uncertainty, i.e., the first layer of requirements (as introduced in \Cref{sec:introduction}), reducing the need for later adjustments and improving alignment between project ambitions and achievable outcomes.

\subsection{Framework performance and challenges}
We notice that while the framework can be applied, the application demands certain level of familiarity with the available source material. Not every project is composed in the same way and no universal template exists. As we have seen with the case of PrimaVera, we had to rely on our interpretation of the targeted TRL levels, as TRL scale was not used in the proposal. There is room to adjust and streamline the flow of our framework, which will require going through more project cases.

Generally speaking, achieving high demonstration levels (see \Cref{tab:demonstration_levels}) is not trivial. It was mentioned that solutions intended for modern industrial CPS rely on software algorithms and data. Given that what crosses the boundaries of one WP to another is either original data, or artefacts generated from it, a clear understanding of these items is a must. Any level of demonstration covering integration, will require input/output data structures and formatting to be well-defined and covered in the granular requirements. This is expected for all WPs involved in integrations, when demonstration levels~2, 4, and 5 are deemed achievable through our framework.

Collection of results for the presented framework is a challenge in itself, affecting the validation of its effectiveness. Projects in our current portfolio that started prior to this paper have progressed beyond the point of this framework's utilisation. Applying it to projects in early stages will only produce results in 3--4 years from the time of this writing. Nonetheless, we have provided analyses of two projects at such different stages of maturity in \Cref{sec:discussion}.

\section{Related work}
\label{sec:related_work}
The literature often mentions terms such as \enquote{design demonstrator}, \enquote{boundary object}~\cite{Vilsmaier:2015:CMLS, Feldhoff:2019:BTPB}, or \enquote{intermediary objects}~\cite{Vinck:2003:EEED, Vinck:2012:AMCF}. This is partly inspired by the industry practice of new technology development and we see a \enquote{market} or \enquote{commercialisation} stage at the end. None of these term align well with our understanding from a demonstrator, as the final disseminating outcome of a research project. Most literature consider such terms as vessels of transferring artefacts between practitioners, e.g., designers and scientists.

Moultrie~\cite{Moultrie:2015:UCRD} recites that in commercial setups, early involvement of industrial design has proven advantageous, leading to technologies with better applications~\cite{Gemser:2001:HIID, Hertenstein:2001:VDEC}. While~\cite{Moultrie:2015:UCRD} relies primarily on case studies to showcase the categorisation, which is a repeating trend in other literature~\cite{Gopinath:2018:DSRI, Bobbe:2023:DCHD}, the analysis itself comes across as ad hoc. A plausible reason has to be the broadness of classification. That is one motivation for us to keep the domain of applicable research projects fairly limited for the time being.

Considering the aforementioned resources, we do not see any demonstrator typology or requirements elaboration method that is tailored towards multi-stakeholder research projects. In a multi-stakeholder and multi-WP project, use-cases and WP relationships have to be taken into account. As such, our method is designed with and driven by engineering awareness.

\section{Conclusion}
\label{sec:conclusion}
Ambiguities surrounding the topic of \enquote{demonstrators} is a real cause for inefficiency in collaborative research projects. We offer a practical typology that helps align stakeholder expectations and is better suited to research projects, as opposed to the TRL scale. The absence of a reference scale suitable for research projects is noted to be a key obstacle for planning and alignment between stakeholders. The case of PrimaVera is a clear example.

Additionally, we presented a framework designed to tackle this challenge in a systematic fashion. Our framework relies solely on information and documentation available at the time of proposal composition. As such, misalignments and missing detail crucial to the formulation of demonstrators could be revealed early on. In fact, the primary value of the framework lies in supporting demonstrator definition at the project-level, i.e., the first layer of requirements, where adjustments can be applied almost effortlessly.

We have shown in \Cref{subsec:the_case_of_zorro} how this framework can be applied to the project ZORRO (a project in its early stages). We have also shown in \Cref{subsec:the_case_of_primavera} the issues revealed by our framework in the project PrimaVera (a project now nearly complete). Had the framework been applied to PrimaVera during proposal composition, practical challenges encountered in the project's demonstrators could have been prevented.

Two extensions can be considered to this paper as potential immediate future work. First, full or partial automation of the proposed framework through a dedicated tool, supported by well-defined artefact formats and input templates. Such a tool would enable systematic execution of the framework. WP dependencies and related interactions can be formally represented as Directed Acyclic Graphs (DAGs), allowing algorithmic traversal, constraint propagation, and automated detection of bottlenecks. Second, an extensive look into artefact quality, i.e., code and data quality, is in demand, as the notion of quality is context-dependent and can be interpreted in numerous ways. While we have limited the usage of our method to software-intensive systems, applying it to more diverse use-cases could prove successful.

Finally, we see potential in extending our framework beyond proposal composition, supporting continuous tracking and monitoring of WP progress and demonstrator feasibility during the project lifecycle. By detecting realised risks early on, planning could be adjusted.

\begin{acks}
This publication is part of the project ZORRO with project number KICH1.ST02.21.003 of the research programme Key Enabling Technologies (KIC), which is (partly) financed by the Dutch Research Council (NWO).
\end{acks}


\balance

\printbibliography

\end{document}